# Complex Ferromagnetism and Magnetocaloric Effect in Single Crystalline $Nd_{0.7}Sr_{0.3}MnO_3$

**R. Venkatesh[1], M. Pattabiraman[1*], S. Angappane[1†], G. Rangarajan[1], K. Sethupathi[1], Jessy karatha[1,2], M. Fecioru-Morariu[2], R.M. Ghadimi[2] and G. Guntherodt[2]**

[1] *Department of Physics, Indian Institute of Technology, Madras, Chennai 600 036, India.*

[2] *Physikalisches Institut, RWTH Aachen, 52056 Aachen, Germany*

Abstract

The magnetocaloric effect in single crystalline $Nd_{0.7}Sr_{0.3}MnO_3$ is investigated by measuring the field - induced adiabatic change in temperature ($\Delta T_{ad}$) which reveals a single negative peak around 130 K well below the Curie temperature ($T_C$ = 203 K). In order to understand this unusual magnetocaloric effect we invoke the reported $^{55}$Mn spin-echo nuclear magnetic resonance, electron magnetic resonance and polarized Raman scattering measurements on $Nd_{0.7}Sr_{0.3}MnO_3$. We show that this effect is a manifestation of a competition between the double exchange mechanism and correlations arising from coupled spin and lattice degrees of freedom which results in a complex ferromagnetic state. The critical behavior of $Nd_{0.7}Sr_{0.3}MnO_3$ near Curie temperature is investigated to study the influence of the coupled degrees of freedom. We find a complicated behaviour at low fields in which the order of the transition could not be fixed and a second order like behaviour at high fields.

**PACS: 75.30.Sg, 75.47.Lx**

**Keywords: Magnetocaloric effect, single crystal, manganites**

---

* Corresponding Author: pattu@physics.iitm.ac.in

† Present Address: Department of physics, Sungkyunkwan University, Suwon 440-746, South Korea.

# 1 INTRODUCTION

Doped manganites with the general formula, $R_{1-x}A_xMnO_3$ (R= La, Nd, Pr etc.; A= Ca, Sr etc.) exhibit a significant magnetocaloric effect (MCE) (i.e.) external magnetic field induced temperature change in addition to colossal magnetoresistance. A complex interplay among spin, lattice, charge and orbital degrees of freedom dictates the physics of manganites and is likely to influence the magnetocaloric properties. However, magnetocaloric studies in manganites always focus on a temperature window centered about the Curie temperature ($T_C$), where a peak is observed in the isothermal entropy and adiabatic temperature change ($\Delta T_{ad}$) so as to consider the feasibility of using these compounds as magnetic refrigerants at relatively high temperatures [1, 2].

In this study we report, for the first time, an unusual magnetocaloric effect observed in single crystalline $Nd_{0.7}Sr_{0.3}MnO_3$ (NSMO-0.3) where a negative peak in $\Delta T_{ad}$ is observed well below the Curie temperature (203 K) in the ferromagnetic metallic state. Combining this result with our previous $^{55}$Mn spin echo Nuclear Magnetic Resonance (NMR), polarized Raman scattering and Electron Spin Resonance (ESR) studies on NSMO-0.3, we show that this effect is a manifestation of the interplay between spin and lattice degrees of freedom resulting in a complex ferromagnetic state. It is seen that the direct measurement of the magnetocaloric effect is a useful tool to study the temperature evolution of this interplay in manganites.

The nature of the paramagnetic to ferromagnetic transition in NSMO-0.3 and the influence of the interplay among various degrees of freedom are further investigated by an analysis of critical behaviour observed near $T_C$ at low and high magnetic fields.

## 2 EXPERIMENTAL DETAILS

Single crystals of $Nd_{0.7}Sr_{0.3}MnO_3$ used in this work were grown in an infrared image furnace by the floating zone technique [3]. The ESR, Raman and MCE measurements were made on the same batch of samples, all of which were cut from one large single crystal of high quality. The crystal was characterized by x-ray diffraction, electrical resistivity, ac susceptibility and dc magnetization measurements. The magnetization and MCE measurements were made on a non-oriented spherical sample. The demagnetization field is calculated to be 21 Oersted. Its effect is negligible in critical exponent analysis as the slope change in the modified Arrott plot (see below) is less than 0.1%.

Single crystal XRD data was collected using an Enraf - Nonius CAD-4 diffractometer. A small NSMO-0.3 crystal of dimensions 0.4 x 0.25 x 0.175 mm was used for data collection. The figure 1(a) shows the unit cell of $Nd_{0.7}Sr_{0.3}MnO_3$ showing the various atomic coordinates. The unit cell consists of four distorted cubic units with Nd or Sr atoms occupying the corner of the cube, Mn atoms at the body centre and O atoms at the face centre. The crystal system is orthorhombic with space group Pmmm and refined unit cell parameters, $a$ = 7.712(3) Å, $b$ = 7.723(5) Å, $c$ = 7.727(7) Å. The sum of the occupancies of Nd in unit cell is constrained to 7.5(~70%) and Sr to 2.5(~30%). More details may be found in [4].

AC magnetic susceptibility measurements were carried out using a commercial ac susceptometer (Sumitomo, Japan) at an operating frequency of 313Hz and an applied ac magnetic field with r. m. s. amplitude of 0.1 Oersted. Electrical resistivity measurements were carried out using the four-probe method in the temperature range from 50 K to room temperature in a Janis closed cycle refrigerator. The concomitant metal-insulator and paramagnetic-ferromagnetic transitions observed are shown in fig. 1(b). Magnetization data used for critical exponent analysis were measured using an MPMS Quantum Design SQUID magnetometer.

### *2. 1 Magnetocaloric effect*

Temperature changes induced by adiabatically switching off an external magnetic field can be measured in two ways. (a) Indirectly, by estimating the isothermal change in entropy (ΔS) and (b) directly, by measuring the field induced adiabatic temperature change $\Delta T_{ad}$. We have adopted both the methods to measure the magnetocaloric effect.

*Indirect MCE measurement:* The entropy change due to variation of the magnetic field [2] from 0 to $H_{max}$ can be calculated from the Maxwell's relation using

$$\Delta S_H = \int_0^{H_{max}} \left(\frac{\partial M}{\partial T}\right)_H dH \tag{1}$$

Heat capacity (Cp) measurements were then combined with the temperature dependence of entropy change in magnetic field ($\Delta S_H$) to get the adiabatic temperature change. If we assume that $T/C_P(H)$ varies much slower with H than does $\left(\frac{\partial M}{\partial T}\right)_{H_i}$, which is a good approximation in the paramagnetic-ferromagnetic transition temperature range, the adiabatic temperature change can be written as

$$\Delta T_{ad}\left(\text{indirect}\right) = \frac{T}{C_p(H)} \Delta S_M(H) \tag{2}$$

$C_P$ (H) denotes the specific heat in an applied magnetic field. In our case we have used the zero field heat capacity value instead and are getting a lower limit for the adiabatic temperature change as $C_P$ (H) is a decreasing function of H [5]. The temperature dependence of magnetization data were field cooled measurements done in an EG&G vibrating sample magnetometer and heat capacity measurements using NETZSCH Differential Scanning Calorimeter (DSC 200PC "Phox") by ratio method.

*Direct MCE measurements:* The direct measurement of adiabatic change in temperature was made by mounting the NSMO-0.3 crystal in a Teflon holder with a platinum resistance thermometer in direct contact with the sample. The temperature was monitored continuously using a Lakeshore temperature controller. The influence of the magnetic field on the sensor was tested in advance and found to be negligible. The sample holder was placed inside a liquid nitrogen cryostat between the poles of an electromagnet. Adiabatic conditions were ensured by evacuating the sample space. The sample was cooled down to the lowest temperature and readings were taken in natural warming process at a slow rate of about 0.2 K/min. A field of 1 Tesla was applied for about 30 seconds. The adiabatic temperature change $\left(\Delta T_{ad}\left(\text{direct}\right)\right)$ was defined as the difference between the temperatures recorded before and after applying the magnetic field. The temperature change was measured in steps of 5 K.

# 3 RESULTS

### *3.1. Magnetocaloric Effect*

*Indirect MCE measurement*: The temperature (T) dependence of magnetization (M) measured at different magnetic fields for NSMO-0.3 single crystal is shown in the figure 2(a). The Curie temperature is found to be 203 K corresponding to the maximum slope change in the M-T curve in an applied field of 0.1 T. The entropy change associated with magnetic field variations calculated from eq. 1 is shown in fig. 2(b). The field induced adiabatic temperature change $\left(\Delta T_{ad}\left(\text{indirect}\right)\right)$ computed from zero field heat capacity using equation (2) is shown in fig. 3. The inset shows the temperature variation of the zero field heat capacity of NSMO-0.3. The indirect measurements of both $\Delta S_H$ and $\left(\Delta T_{ad}\left(\text{indirect}\right)\right)$ shows a maximum around the Curie temperature corresponding to the para –ferromagnetic phase transition.

*Direct MCE measurement*: The temperature dependence of the measured adiabatic temperature change $\left(\Delta T_{ad}\left(\text{direct}\right)\right)$ is shown in fig. 4. A prominent negative peak is observed at 130 K which is well below $T_C$ (203 K). A small positive peak is observed around $T_C$ and is much smaller than the peak computed in $\left(\Delta T_{ad}\left(\text{indirect}\right)\right)$.

**3.2 *Discussion on MCE Measurements***

The observation of a pronounced field - induced temperature change around 130 K, well below $T_C$ is unusual. The observed magnetocaloric effect is negative. That is the temperature decreases upon application of a field. Such a large negative MCE peak below $T_C$ is unique and to the best of our knowledge has not been reported before in manganites. In order to understand the cause of this unusual MCE, we turn to our previous results on the temperature dependence of zero field $^{55}$Mn spin-echo NMR [6], polarized Raman scattering [7] and electron magnetic resonance (EMR) on $Nd_{0.7}Sr_{0.3}MnO_3$ [8]. It turns out that there are distinct changes in several physical quantities measured by these diverse microprobes around the same temperature range where the MCE peak is observed. This is discussed below in detail.

*(1) Zero Field $^{55}$Mn spin-echo NMR*: The $^{55}$Mn NMR line shape of NSMO-0.3 is multi-peaked throughout the temperature range of measurement [6]. When the centre of gravity of the lineshape [see note in ref 9] is plotted as a function of temperature it is seen that the critical behaviour seen by NSMO-0.3 is significantly ‘slower’ than that expected from the mean-field prediction of a Heisenberg ferromagnet (solid line in fig. 5(a)). Moreover there is a distinct kink around 130 K. This is due to an abrupt weight shift in the lineshape towards higher frequencies (implying higher local (electronic) magnetization) and is suggestive of enhanced ferromagnetic order below 130 K.

*(2)Polarized Raman Scattering:* The Raman spectrum in the XX polarization of single crystalline NSMO-0.3 reveals several phonon bands up to 1000 $cm^{-1}$ [7]. Among these the phonon peak at 65 $cm^{-1}$ ($M_{rot}$) is a rotational mode and is superimposed on a low frequency response from mobile carriers due to strong (collision dominated) electronic scattering from spin fluctuations, local lattice distortions etc. and is usually observed in a ‘dirty’ metal with a high scattering rate.

In manganites such as NSMO-0.3 which exhibit a paramagnetic-insulator to ferromagnetic-metal transition upon cooling, the electronic scattering response reduces significantly below $T_C$. In NSMO-0.3 the electronic scattering response decreases abruptly around 130 K resulting in a prominent narrowing of mode $M_{rot}$ (fig. 5(b): left axis). This suggests a sudden decrease in electronic scattering from spin fluctuations and/or local lattice distortions.

Another mode around 491$cm^{-1}$ ($M_{JT}$) corresponds to out-of-phase stretching vibrations of the $MnO_6$ octahedra and is associated with the Jahn-Teller distortion of the octahedra. $M_{JT}$ exhibits an anomalous softening around 150 K (fig. 5(b): right axis) pointing to the presence of a substantial JT distortion below $T_C$ which weakens below 150 K.

There appears to be a striking correspondence between the temperature evolution of the internal distortions of the $MnO_6$ octahedra (as probed by Raman scattering) and microscopic (electronic) magnetism (as probed by $^{55}$Mn NMR). This suggests that the local lattice distortions may be coupled to the spin degrees of freedom.

Electron magnetic resonance measurements correlated with reported synchrotron X-ray scattering lend support to this suggestion. In a perfect crystal only Bragg peaks are observed from X-ray or neutron scattering. In the presence of polaronic lattice distortions there are small deviations from the perfect crystalline structure. These deviations induce a finite X-ray or neutron scattering intensity close to a Bragg peak and can be seen from synchrotron X-ray scattering. For Jahn-Teller polarons, this extra intensity has a butterfly shape in momentum space. In addition to this diffuse polaron scattering, satellite peaks located at charge arrangements corresponding to the CE (antiferromagnetic spin, charge and orbital order) state as observed in half-doped manganites have been found in NSMO-0.3 [10]. The intensity of this satellite peak which is indicative of correlated lattice distortions with short range change and/or orbital order (COO) goes through a maximum as NSMO-0.3 is cooled through $T_C$ as shown in fig. 6. The peak intensity is still significant below $T_C$. EMR spectra for NSMO-0.3 obtained above and below $T_C$ are shown as insets in the same figure. The EMR spectra at 210 K and 300 K correspond to the paramagnetic phase as $T_C$ is ~ 203 K. However it is seen that this paramagnetic spectrum persists down to 190 K and ferromagnetic resonance is observed only below 180 K. This indicates that long-range ferromagnetic order does not set immediately below $T_C$.

*(3) Electron Magnetic Resonance (EMR):* The EMR spectrum below 180 K is complex and difficult to analyze. When the centre of gravity of the spectrum [See note in Ref. 9] is plotted as a function of temperature (fig. 5(c)) it is seen that above 140 K the spectrum is noisy with no systematic temperature dependence. Below 140 K the centre of gravity increases smoothly suggesting an enhanced ferromagnetic spin order. In the paramagnetic state the EMR measurements on NSMO-0.3 provide strong evidence for the presence of correlations between spin and lattice degrees of freedom. In several manganites the paramagnetic linewidth ($\Delta H$) exhibits 'quasi-linear' temperature dependence due to spin-spin interactions. When appropriately normalized, $\Delta H$ vs T for most manganites lie on the same 'universal curve'. It has been observed that for NSMO-0.3 the line width exhibits a marked deviation from the universal curve and that $\Delta H$ vs T is linear above $T_C$ [11]. This behaviour has been interpreted in terms of a spin-phonon interaction in which the spins are strongly coupled with lattice vibrations.

*(4) Magnetocaloric effect:* The direct MCE measurements made in the present study is re-plotted in (fig. 5(d)). Significant changes in the four different measurements described above occur within a small temperature window around 130 K as shown in fig. 5. NMR, Raman, EMR and synchrotron X-ray scattering studies suggest that correlated entities (clusters) with coupled spin-lattice degrees of freedom exist above and below $T_C$ in NSMO-0.3 and are likely to compete with the double exchange mechanism responsible for ferromagnetism in manganites. The correlations among these diverse measurements suggest that the coupling between the spin and lattice degrees of freedom abruptly reduces below 130 K and the associated change in entropy has resulted in a significant MCE.

The large $\Delta T_{ad}$ observed around 130K has a negative sign. Thus the entropy change occurring across 130 K does not arise from ferromagnetic spins (since they result in a positive $\Delta T_{ad}$). It has been recently shown that the magnetocaloric contribution from charge and orbital order alone has a positive $\Delta S$ contribution [12] and hence the resulting $\Delta T_{ad}$ has a negative sign. The disruption of charge and orbital order in presence of a magnetic field results in an increase in entropy and decrease in temperature. We suggest that the observed MCE about 130K is associated with the disruption of the correlated COO entities. Thus, the sign and magnitude of the 'direct' MCE peak is singular evidence pointing to the nature and robustness of the competition between the mechanisms responsible for double exchange and interplay among spin, lattice and possibly orbital degrees of freedom. The small positive $\Delta T_{ad}(\text{direct})$ observed across $T_C$ may also be interpreted in terms of this competition. Across $T_C$ the total adiabatic change in temperature has two contributions one due to ferromagnetism $\Delta T(\text{FM})$ and another due to charge/orbital ordered $\Delta T(\text{COO})$. Since both contributions have opposite signs the net observed $(\Delta T_{ad})$ is suppressed. The sign of the observed $(\Delta T_{ad})$ suggests that the ferromagnetic component is slightly higher than that due to the COO entities around $T_C$. Equation (2) which assumes the presence of a pure ferromagnetic phase overestimates $\Delta T(\text{indirect})$ (fig. 3) resulting in a large difference with $\Delta T(\text{direct})$ (fig. 4). Such differences in features between computed and measured MCE data has been reported earlier in manganites [13].

Thus it is seen that the complex interplay among spin, lattice and possibly even charge and orbital degrees of freedom strongly influences the magnetocaloric effect in NSMO-0.3. The direct measurement of the magnetocaloric effect is a useful tool to study the temperature evolution of this interplay in manganites.

### 3.3. *Critical behaviour of NSMO-0.3*

In an attempt to analyze the nature of ferromagnetism in NSMO-0.3 quantitatively, we turned to an analysis of the critical behaviour near $T_C$. DC Magnetization (M) data taken up to 2.1 T around the Critical temperature (203 K) with 1 K intervals (fig. 7(a)) was used for the analysis. According to the criterion proposed by Banerjee [14], a magnetic transition is of first order (second order) if the slope of the plot between $M^2$ and H/M is negative (positive). In our case as shown in fig. 7(b) the slopes for all isotherms up to 206 K are positive. Isotherms above 206 K exhibit a small negative slope at low fields and a positive slope at high fields.

Earlier reports on manganites suggest that Magnetization alone cannot determine the order of the transition and that the interplay among spin, lattice orbital and charge degrees of freedom should be taken into account [15,16,17]. We show that this observation is also applicable to NSMO-0.3 by invoking the existence of a Griffith's phase in the paramagnetic state just above $T_C$.

This paradigm presupposes the suppression of ferromagnetic transition due to disorder [18]. There exists a Griffith's temperature $T_G$ ($>T_C$) at which the transition would have occurred in the absence of disorder. The temperature window between $T_C$ and $T_G$ is regarded as the Griffiths phase in which a disorder dependent distribution of exchange energies (and therefore $T_C$' s) exist. We found a sharp downturn in the inverse of the paramagnetic magnetization $\chi(T)$ above $T_C$. This can be well explained by the existence of a Griffith's temperature at 207 K. $T_G$ is obtained by a fit to a power law behavior [18], $\chi^{-1}(T) \propto (T - T_G)^{1-\lambda}$, with $\lambda = 0.45$.

We now relate this Griffith's temperature to the magnetization data. According to the Landau theory of phase transitions, the Gibbs free energy can be written as,

$$G(T,M) = G_o + \frac{1}{2}A(T - T_o)M^2 + \frac{1}{4}BM^4 + \frac{1}{6}CM^6 - MH \tag{3}$$

with A, C > 0. From minimization of G (T, M)

$$\frac{H}{M} = A(T - T_o) + BM^2 + CM^4 \tag{4}$$

As mentioned above the condition for first order is B < 0 where B corresponds to initial slope of H/M vs. $M^2$ isotherms (Arrott plot). For a first order transition a maximum is observed in d M /d H while this is not the case for a second order transition (B > 0).

For NSMO-0.3 (fig. 7(c)) it is seen that that the initial slope observed in the Arrott plot is negative above $T_G$ = 207 K and positive below $T_G$. Correspondingly a maximum in dM/dH is observed above $T_G$ = 207 K and not below. This is quite unconventional and not expected in a system exhibiting a first or second order phase transition. Thus the transition cannot be declared as first order on the basis of the Banerjee criterion in NSMO-0.3. The correlations shown in fig. 5 and corresponding arguments based on NMR, Raman and ESR measurements suggest that the 'disorder' responsible for the Griffiths phase is related to the competition between ferromagnetism and mechanisms behind the COO clusters. Thus the magnetic origin is coupled to the lattice, charge and orbital degrees of freedom.

Considering the rather complicated magnetic behaviour observed and in the absence of a concrete theory of critical exponents when the magnetization is related to several degrees of freedom it is difficult to fix the order of transition in NSMO-0.3.

We may expect the situation to be simpler at high fields when the effect of charge/lattice and orbital degrees of freedom are suppressed in a ferromagnet and the order parameter can be identified with the macroscopic magnetization. Following this assumption we now proceed to determine the critical exponents for NSMO-0.3 at high fields. Since only positive slopes are observed at high fields in all Arrott plot isotherms this calculation assumes the presence of a second order like transition. We later check the validity of our assumptions using the computed critical exponents.

The FM – PM transition is characterized by a set of critical exponents in the critical region, $\beta$ (associated with the spontaneous magnetization ($M_s$)), $\gamma$ (associated with the initial susceptibility ($\chi_o$), and $\delta$ (describes the magnetisation dependence on the magnetic field (H) at $T_C$). They are defined as

$$M_S = m_o |t|^{\beta} \qquad t \geq 0 \tag{5}$$

$$\chi_s^{-1} = h_o/m_o |t|^{\gamma} \quad t \leq 0 \tag{6}$$

$$H = D M^{\delta} \qquad t = 0 \tag{7}$$

Where t is reduced temperature ($1-T/T_C$), $m_o$, $h_o$, and D are critical amplitudes.

Although critical exponents can be determined independently, they are related to each other by scaling equations which is taken in most cases to be of form

$$\beta + \gamma = \beta\delta \tag{8}$$

Since the Arrott isotherms are linear only at high fields to accurately determine the Curie temperature as well as the critical exponents β, γ, and δ the modified Arrott plot technique is used. Modified Arrott plot is used for determining critical exponents which has a scaling equation of state

$$(H/M)^{\frac{1}{\gamma}} = at + bM^{\frac{1}{\beta}} \tag{9}$$

This causes isothermal curves of M(H) data to fall into a set of parallel straight lines in a plot of $M^{1/\beta}$ vs $(H/M)^{1/\gamma}$ if the correct values of β and γ are chosen [19]. The intercepts of the isotherms on the *x* and *y* axes are $(1/\chi)^{1/\gamma}$ for $t > 0$ and $M_S^{1/\beta}$ for $t < 0$, respectively. The isothermal line that passes through the origin is the critical isotherm at $T=T_C$. To find the correct values of β and γ an initial choice of β and γ is made, yielding quasi-straight lines in the modified Arrott plot. From these initial values linear fits to the isotherms are made to get the intercepts giving $M_S(T)$ and $\chi_o(T)$, and an initial value of $T_C$ is determined from the isotherm that passes through the origin. A new value of β is obtained from a ln ($M_S$) vs ln (t) plot by fitting the data to a straight line, the slope of which gives β. Similarly a new value of γ is obtained from ln(1/χ) vs ln(t) plot. These new values of β and γ are then used to make modified Arrott plots. This procedure is continued till they converge to a stable value. The modified Arrott plot is shown in fig. 8a. only the high field (H > 0.2 T) linear region is used for the analysis (fig. 8 (b)). From the *x* and *y* intercepts in figure 8(b) $M_S$ and 1/χ are measured and shown in figure 8 (c) and 8 (d) respectively. From these the final values obtained are $T_C$=203.6(3) K, β = 0.57(1) and γ=1.16(3).

The M(H) curve at the critical temperature, is shown in fig. 9. From ln-ln plot shown in the inset, the critical exponent $\delta$=3.03(2) is determined. The scaling equation of $\delta=1+\gamma/\beta$ is satisfied in this compound NSMO-0.3. A modified Kouvel-Fischer (KF) [20] analysis is used to independently estimate $T_C$, and $\gamma$. For this, low field ac susceptibility measurements in the critical region are made at a frequency of 313 Hz and a driving ac field of 0.1 Oersted). The KF plot made from real part of the first order susceptibility is used for calculation of the critical exponent as shown in fig. 10. The inverse of the slope gives the value of the susceptibility exponent ($\gamma$) and intercept on the temperature axis is $T_C$ as shown in fig. 10. We have obtained $\gamma$ = 1.17(4) consistent with modified Arrott plot within the error bar and $T_C$ = 203.3 K. These values of critical exponents are between the predicted values for the 3D Heisenberg model ($\beta$ = 0.37 $\gamma$=1.39, $\delta$ = 4.7) and for the mean field model ($\beta$ = 0.5, $\gamma$=1, $\delta$ = 3) as observed in other manganites [21].

The critical exponent analysis can be justified by comparing a plot of M. $t^{-\beta}$ vs H $t^{-(\beta + \gamma)}$ for NSMO-0.3 (fig. 11(a),(b)) and $La_{0.7}Ca_{0.3}MnO_3$ (which exhibits a first order transition) [22]. This plot should ideally have two branches for a second order transition: one above $T_C$ and one below. For $La_{0.7}Ca_{0.3}MnO_3$ such a branching is absent at low and high fields [22] and the observed critical exponents don't match with those predicted by theoretical models or to other manganites. For NSMO-0.3 although the branching is absent for low fields, it is observed at high fields and the critical exponents calculated using the high field data are between those corresponding to mean-field and Heisenberg exponents as observed in several manganites [21].

Moreover for a first order transition modified Arrott plots cannot be used for determination of $T_C$ even by neglecting several low field data points. But For NSMO-0.3 a good estimate of $T_C$ (203.3 K) can be obtained using the high field data justifying our assumption of second order - like behavior at high fields although the critical behaviour is complicated at low fields.

### *3.4 . Discussion on Critical Behaviour of NSMO-0.3*

In what follows we try to explain the field induced change in magnetic behaviour observed in NSMO-0.3. The well known plot between $T_C$ vs tolerance factor (or average A site radius, $\langle r_A \rangle$ ) first reported by Hwang et al for $R_{0.7}A_{0.3}MnO_3$ [23] is shown in fig. 12. The following observations can be made using this plot.

(a) It has been found that compounds with tolerance factor ($t = r_A + r_o / \sqrt{2}(r_B + r_o)$ < 0.92; $r_A, r_B$ and $r_o$ are radii of the A-site, B-site and oxygen ions respectively) exhibit a first order transition with strong spin lattice coupling and volumetric strain effects while those with $t > 0.92$ exhibit a second order transition with a significantly reduced spin lattice coupling [24].

(b) It has recently been found that COO clusters have been observed in several compounds with $t < 0.92$ but not for compounds with t > 0.92 [25].

(c) The idea of randomness of ions occupying the perovskite A site is quantified by the variance $\sigma^2 = \sum y_i r_i^2 - \langle r_A \rangle^2$ where $r_i$ and $y_i$ are the ionic radii and fractional occupancy of the species occupying A site respectively. This was introduced by Rodriguez–Martinez and Attfield [26] who computed the $T_C$ of $R_{0.7}A_{0.3}MnO_3$ manganites when $\sigma^2 = 0$. This curve is also plotted in fig. 12. A striking observation can now be made. For t < 0.92 there is little difference between observed $T_C$ and expected $T_C$ when $\sigma^2$ is zero unlike the case for t>0.92. By combining observations (b) and (c) it is tempting to suggest that A-site disorder and COO are in compatible with each other. When NSMO-0.3 data is also added to fig. 12, it is seen that unlike other compounds with t <0.92 there is a considerable difference between the observed $T_C$ and $T_C$ expected in the absence of A-site disorder.

Compounds exhibiting COO clusters are susceptible to long-range anisotropic strain associated with lattice distortions [27, 25]. The volumetric effects associated with compounds with t <0.92 and the observed first order transition may be linked to this strain. On the other hand it is known that quenched disorder makes a first order transition continuous (or second order) in manganites [28].

As discussed above the critical behaviour of NSMO-0.3 is complex at low fields and second order like at high fields. In order to explain this we propose the presence of a competition between the COO clusters (favouring first order transition) and A-site (quenched) disorder (favouring second order transition). At low fields the competition is robust since the concentration of COO clusters is high. Therefore the observed critical behaviour is complex and the order of the transition is difficult to determine. When the magnetic field is increased, the COO clusters 'dissolve' [10] and the effect of A-site disorder comes into play resulting in a second order like transition.

## 4 CONCLUSIONS

We have studied the influence of correlated clusters on the ferromagnetism of single crystalline $Nd_{0.7}Sr_{0.3}MnO_3$. There is a strong competition between double exchange and the mechanisms responsible for clusters with short range charge/orbital order (COO). It is seen that the complex interplay among spin, lattice and possibly even charge and orbital degrees of freedom strongly influences the magnetocaloric effect in NSMO-0.3. There is a significant reduction in the coupling between spin and lattice degrees of freedom below 130 K resulting in a large negative magnetocaloric response. This competition also results in a suppression of MCE around $T_C$. The direct measurement of the magnetocaloric effect is thus a useful tool to study the temperature evolution of this interplay in manganites. We show that this interplay also affects the observed critical behaviour. We propose a competition between the COO clusters and A-site (quenched) disorder and show that the critical behaviour is complex at low fields with a second order like transition at high fields.

**Acknowledgements**

Authors wish to thank G. Balakrishan, University of Warwick, UK for the single crystal, N. Sivaramkrishnan, SAIF, IIT, Madras for his help with the direct MCE measurements.

**FIGURE CAPTIONS**

**Fig 1**(a) Structural diagram of $Nd_{0.7}Sr_{0.3}MnO_3$ showing the various atomic coordinates. The radius of the circle varies according to the atomic mass (b) Temperature dependent resistivity and real part of ac susceptibility of NSMO-0.3 measured at frequency of 313 Hz and an applied ac field of 0.1 Oe.

**Fig2.** (a) M vs T at different applied magnetic fields; (b) Temperature dependent entropy change of NSMO-0.3.

**Fig 3:** Temperature dependent adiabatic change in temperature; Inset: Temperature dependent zero-field heat capacity of NSMO-0.3.

**Fig 4.** Temperature dependence of adiabatic change in temperature of NSMO-0.3 $\left(\Delta T_{ad}\left(\text{direct}\right)\right)$ in an applied field of 1Tesla. Inset shows the behaviour observed about $T_C$.

**Fig. 5.** Results of $^{55}$Mn spin echo (a) NMR, (b) ESR, (c) Raman scattering and (d) MCE showing changes in the small window from 130K -150K. *(Refer text for details)*

**Fig. 6.** A comparison of ESR spectrum with the structurally correlated synchrotron peak intensity in NSMO-0.3 sample [10]. *(Refer text for details)*.

**Fig 7.** (a) Field dependence of magnetization of NSMO-0.3 at different temperatures; (b) Arrott plot; (c)$1/\chi$ vs T (open circles) in an applied field of 1T. The filled circles denote the temperature corresponding to the M-H isotherms which exhibit a dM/dH peak. The open squares and the filled squares denote the temperatures at which the slope of the Arrott plot is negative and positive respectively. The line is a fit to the power law behaviour expected for a compound with a Griffiths phase.

**Fig. 8**(a) Modified Arrott plot; (b) High field portion of the modified Arrott plot; (c, d) ln-ln plot used to obtain the final values of $\beta$ and $\gamma$.

**Fig 9**. Critical isothermal curve of M(H), taken from at $T_C$ = 203 K . Inset shows ln(M) vs ln(H) plot at $T_C$ ; slope gives $\delta$ = 3.03(2).

**Fig 10.** Modified Kouvel Fischer plots of the real part of the first order susceptibility for the sample $Nd_{0.7}Sr_{0.3}MnO_3$ inverse of slope is $\gamma$=1.17(4) and $T_C$=203.3 K.

**Fig.11** (a) Scaling plot of $M.t^{-\beta}$ vs $H.\ t^{-(\gamma+\beta)}$ for NSMO-0.3 single crystal using $\gamma$ = 1.16 and $\beta$ = 0.57. (b) $M.t^{-\beta}$ vs $H.\ t^{-(\gamma+\beta)}$ plotted in ln-ln scale.

**Fig.12** Open circles denote measured $T_C$ values and the solid line guides the eye (adapted from [23]; Filled circles denote computed $T_C$ in absence of A-site disorder and the dotted line guides the eye (adapted from [26]); The hashed box demarcates the two regions above and below *t* =0.92 (adapted from [24]; *Refer text for details)*. The solid line across the hashed box is a schematic boundary between orthorhombic and Rhombohedral phases (adapted from [25]). PMI: Paramagnetic Insulator; PMM: Paramagnetc Metal; FMM: Ferromagnetic Metal; CO: Charge ordered.

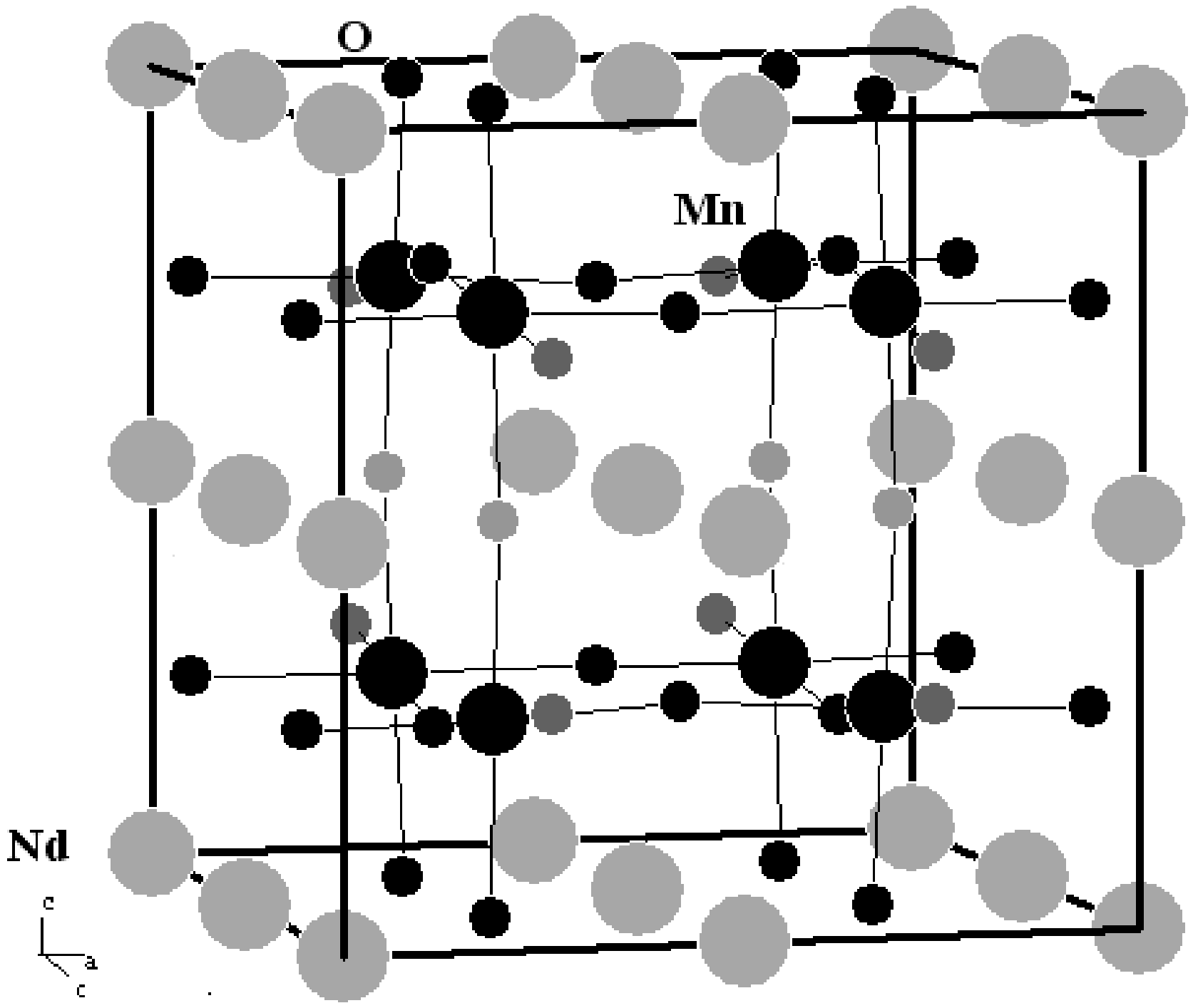

Figure 1(a)

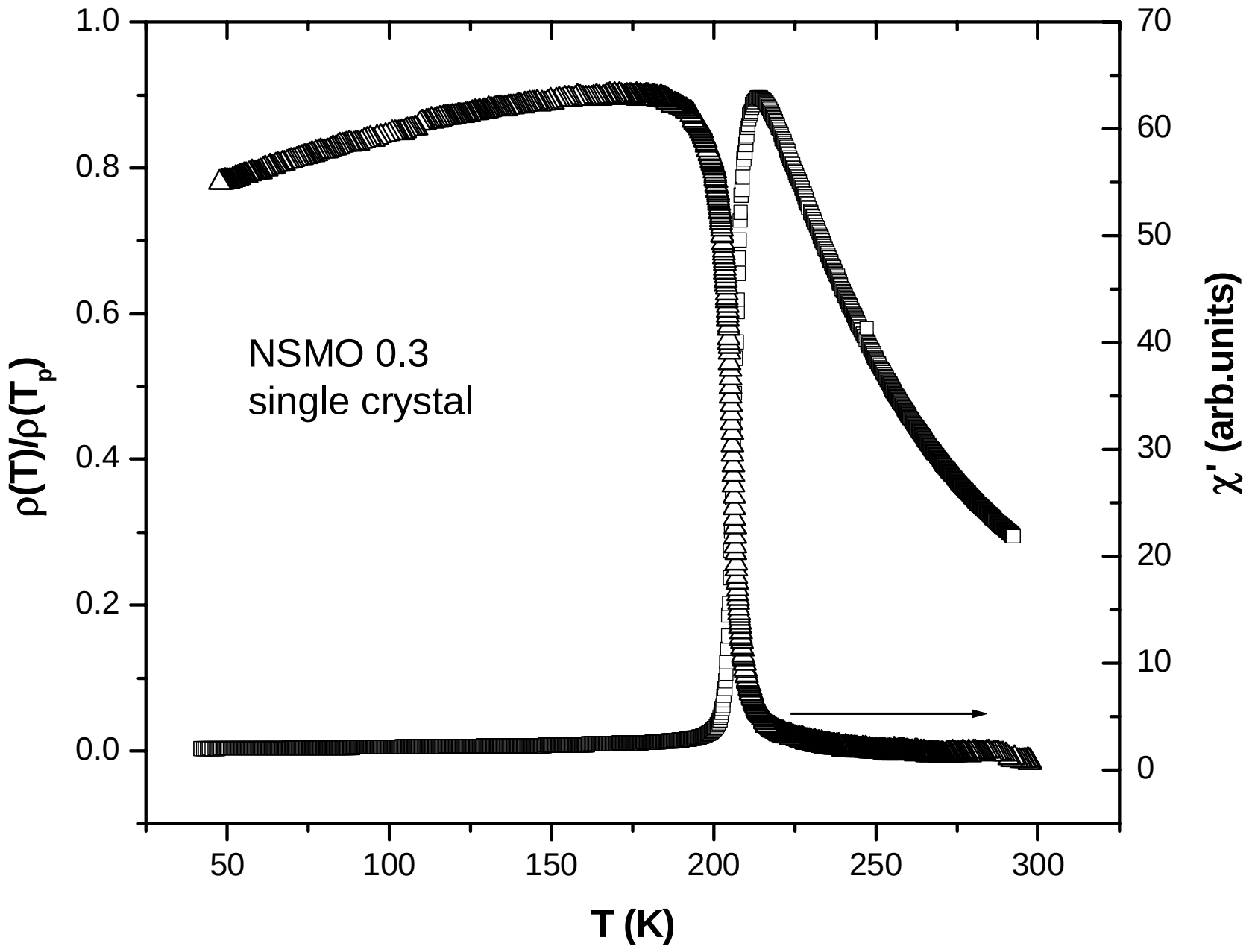

Figure 1(b)

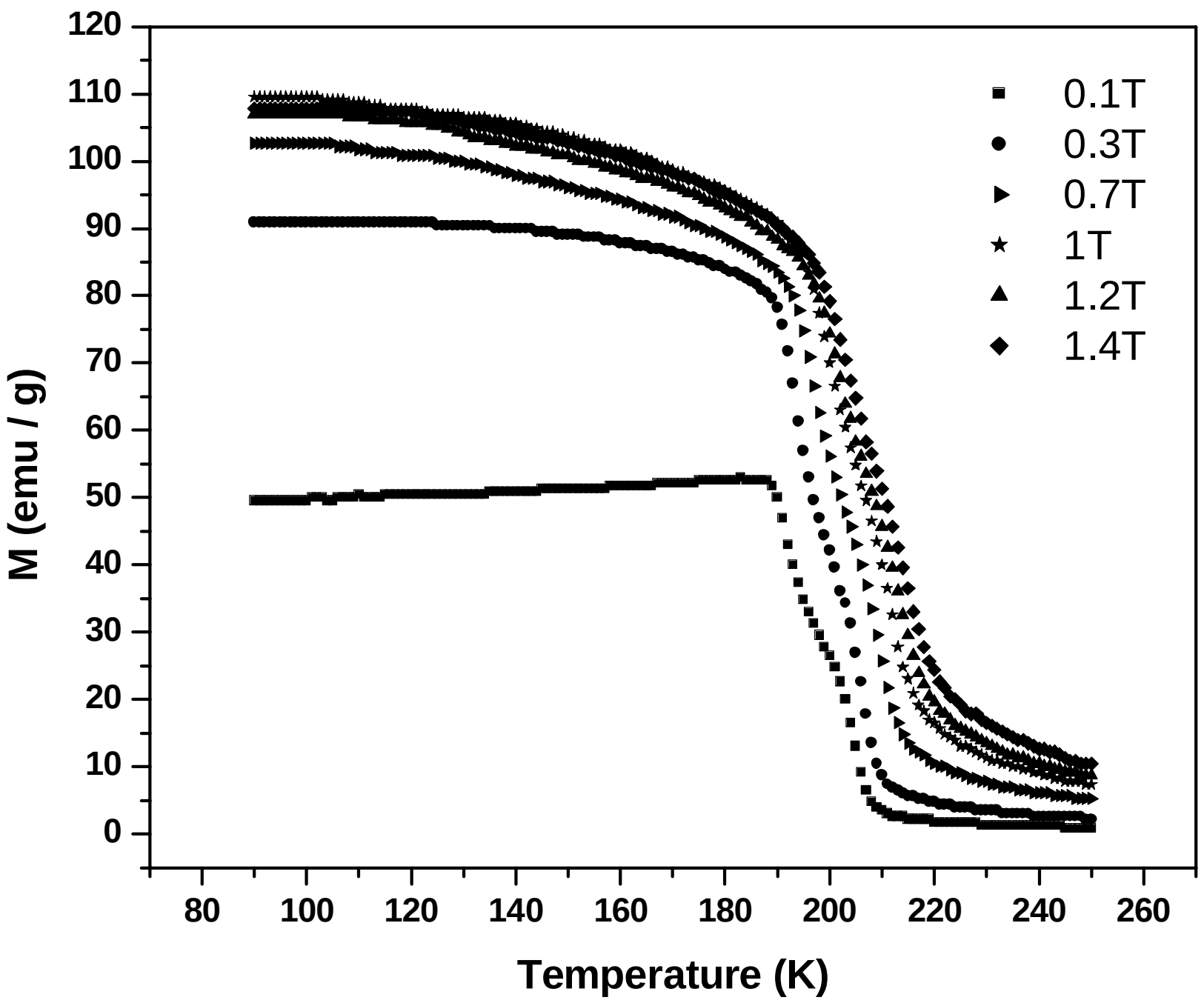

Figure 2(a)

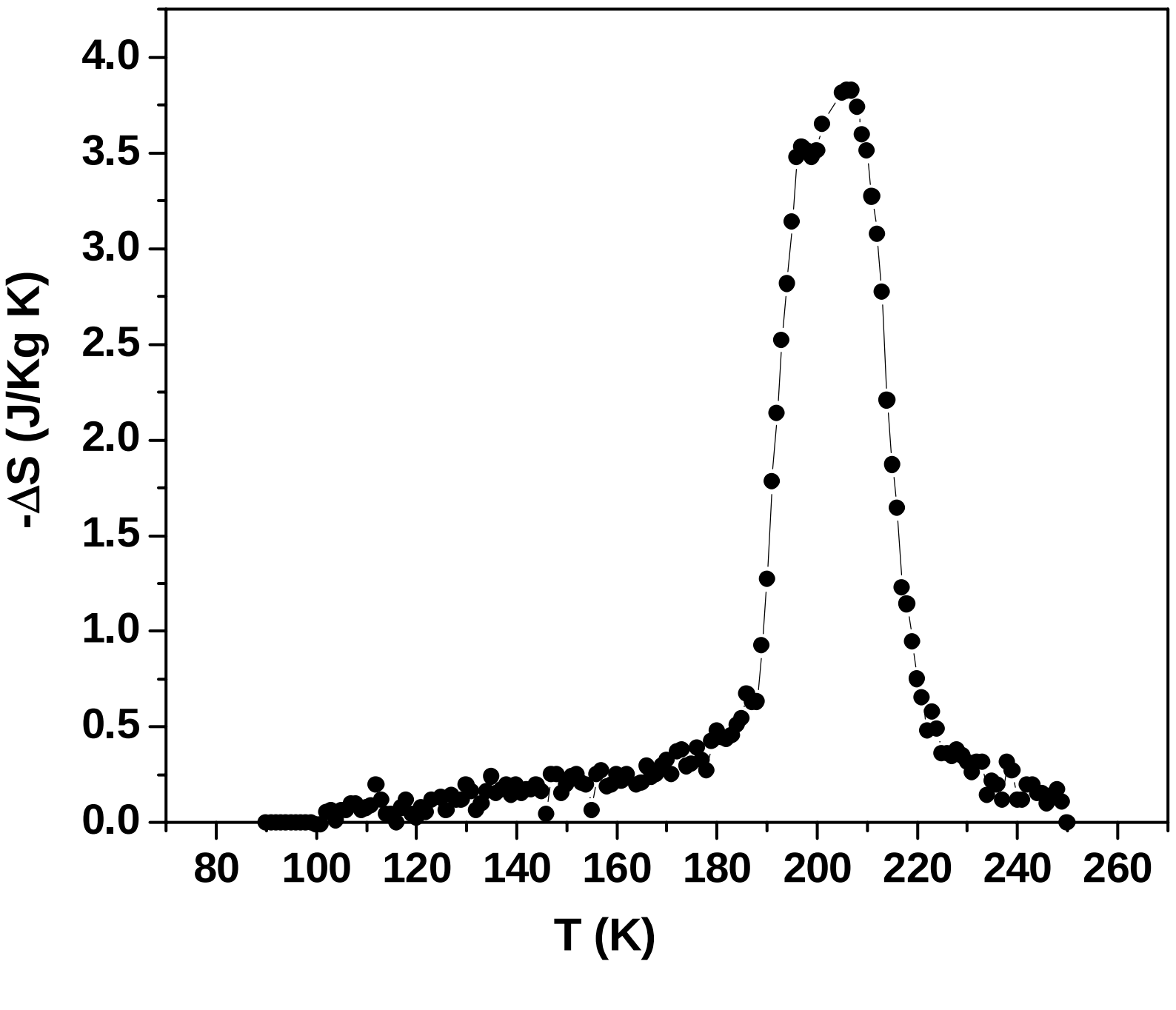

Figure 2(b)

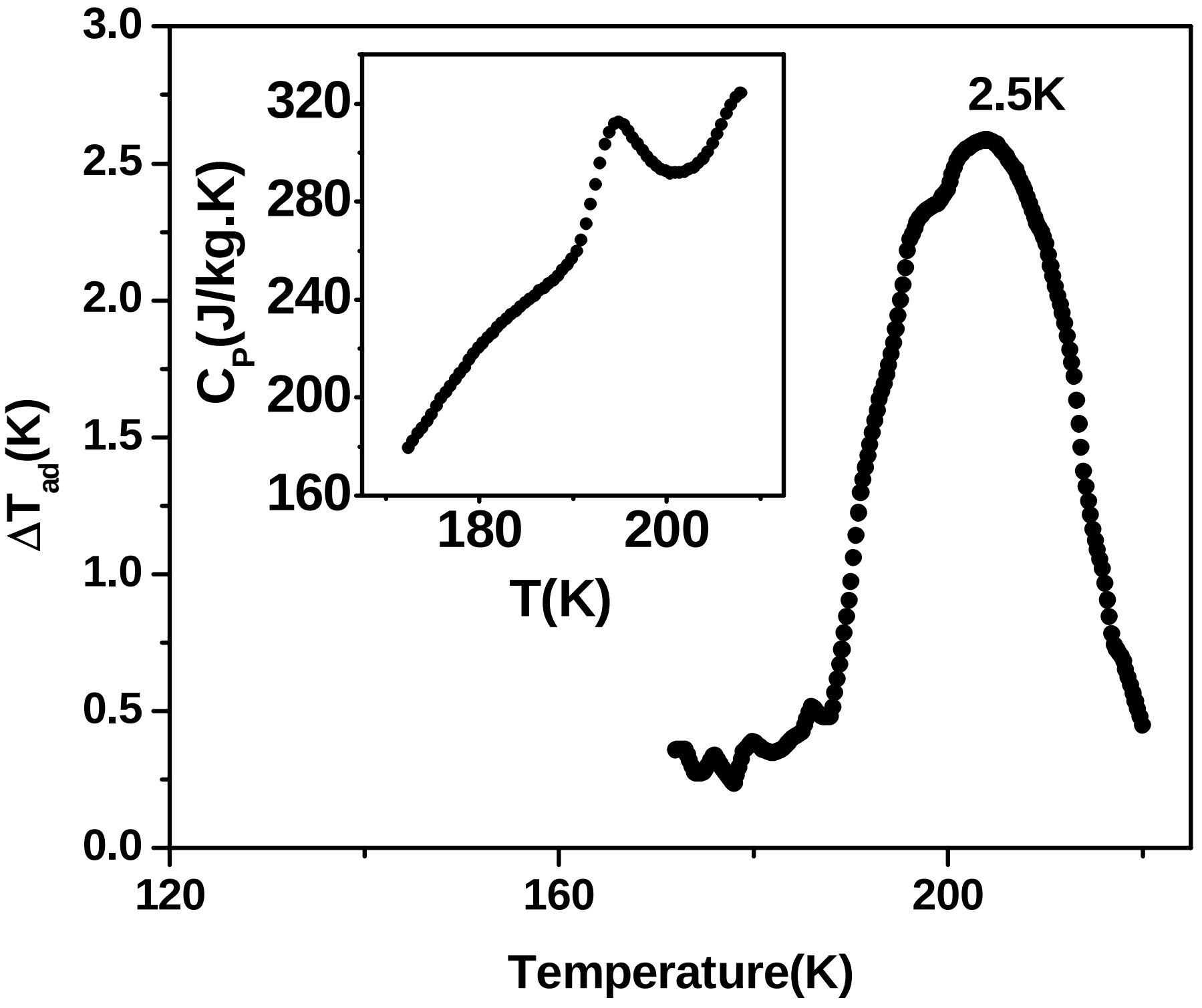

Figure 3

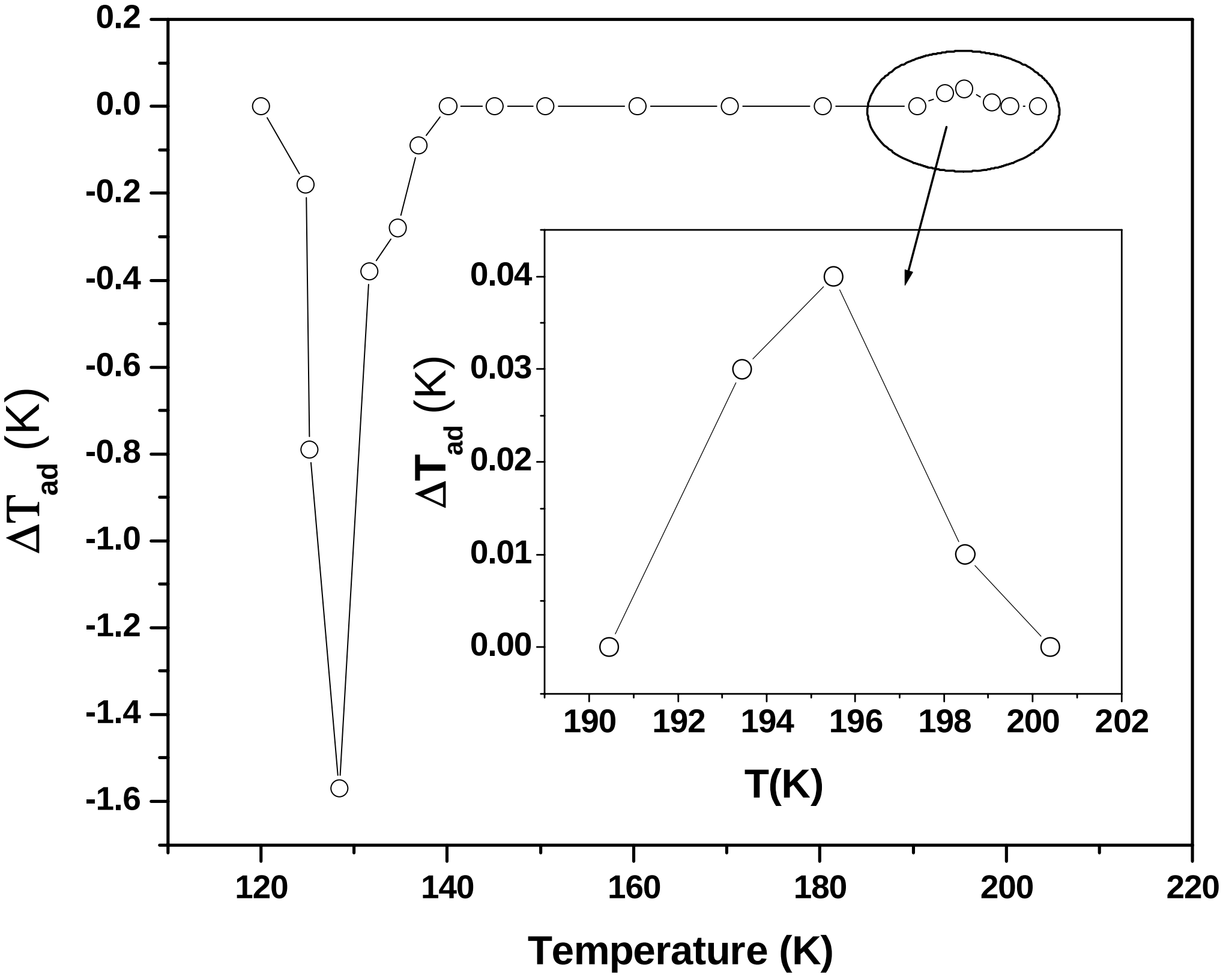

Figure 4

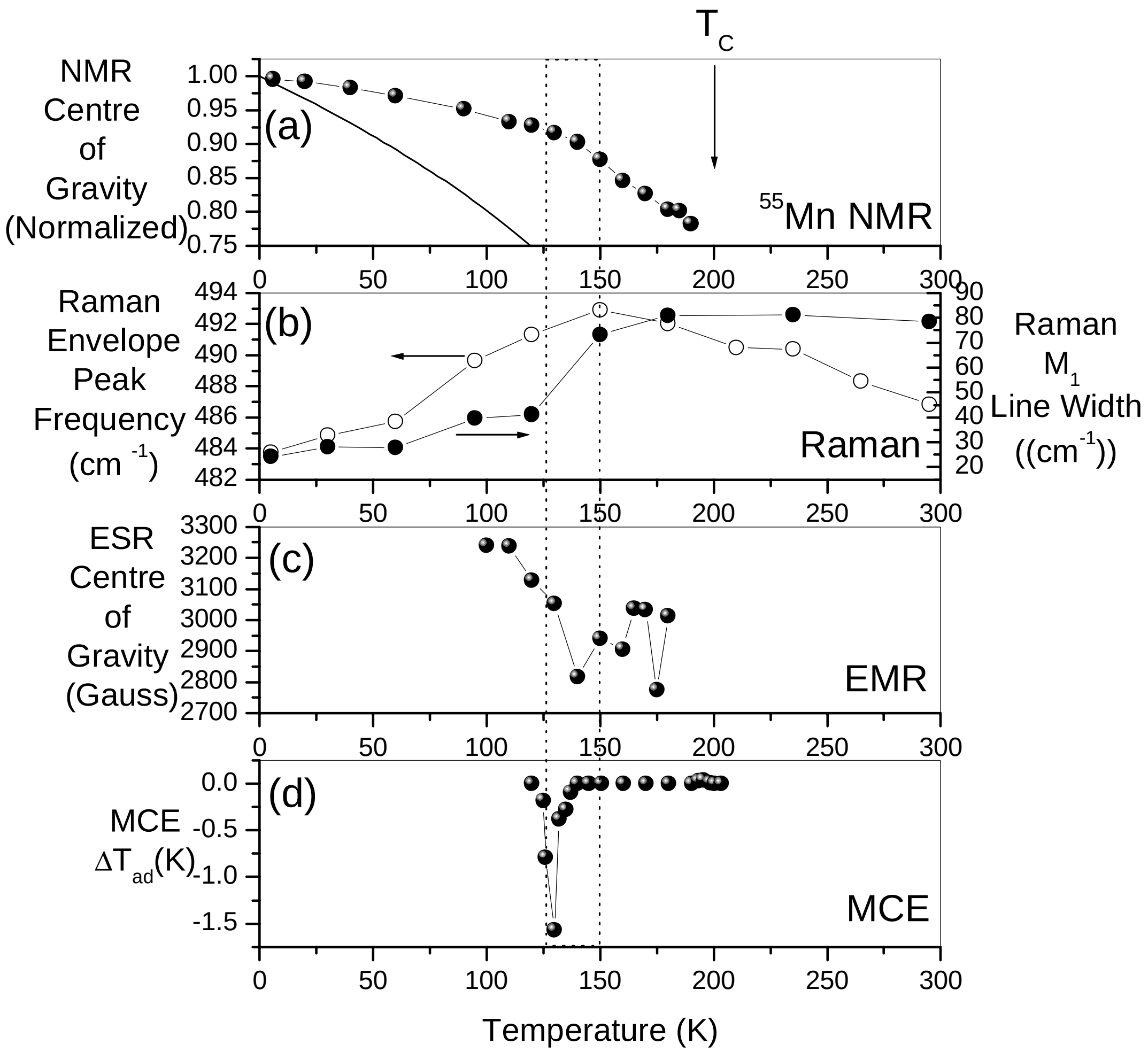

T C
NMR Centre of Gravity (Normalized)
(a)
55Mn NMR
Raman Envelope Peak Frequency (cm -1)
(b)
Raman
Raman M1 Line Width ((cm-1))
ESR Centre of Gravity (Gauss)
(c)
EMR
MCE ΔTad(K)
(d)
MCE
Temperature (K)

Figure 5

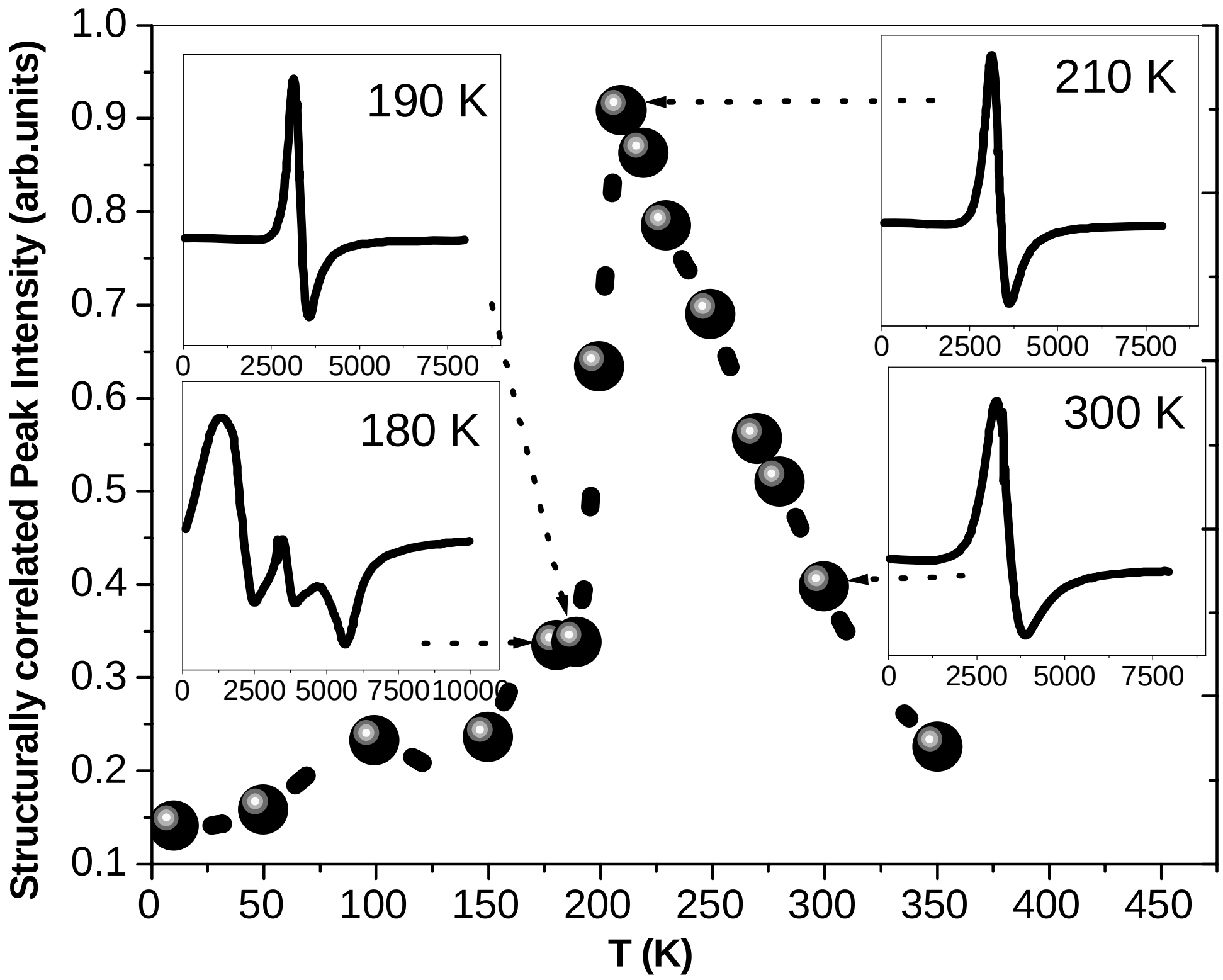

Figure 6

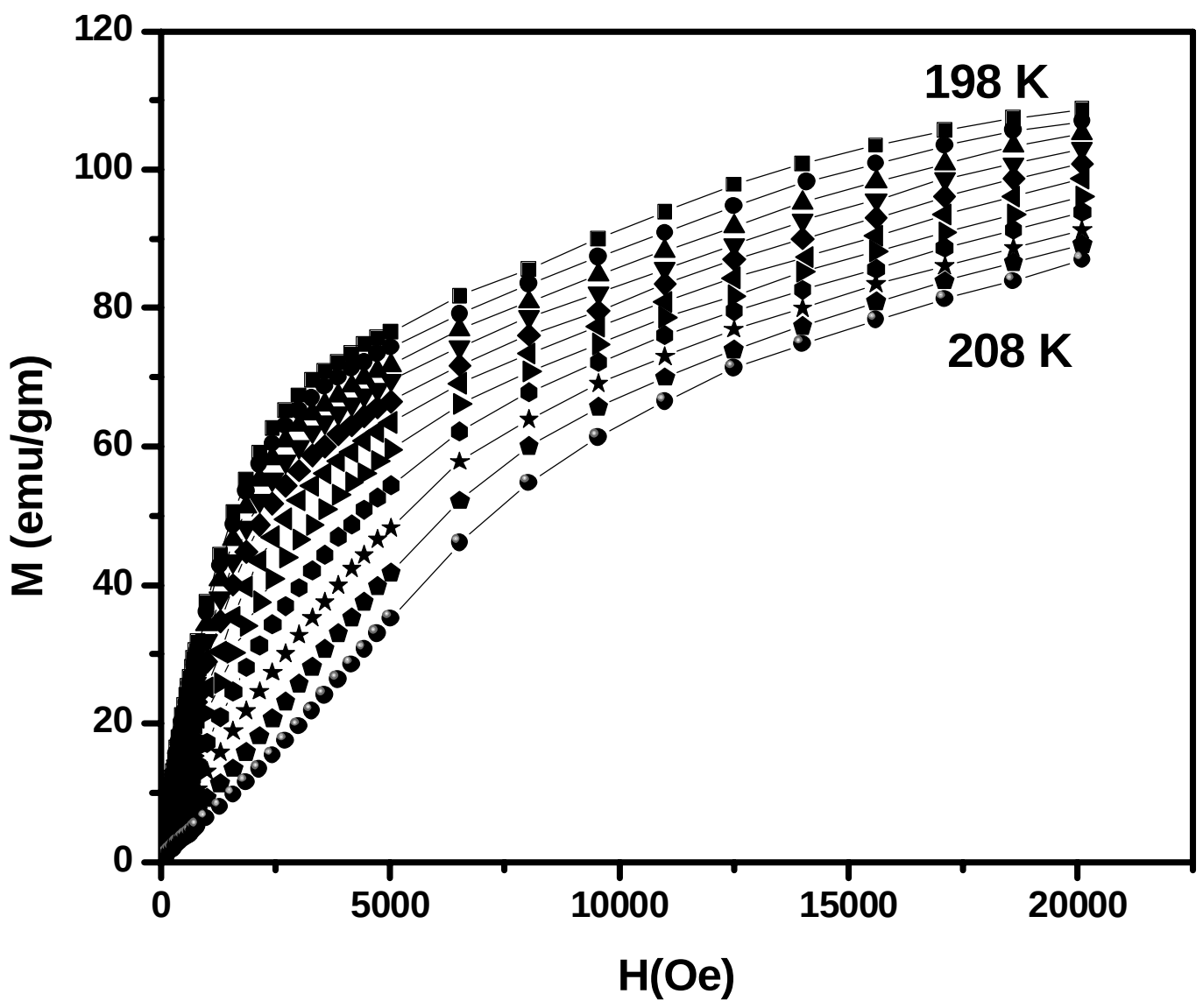

Figure 7 (a)

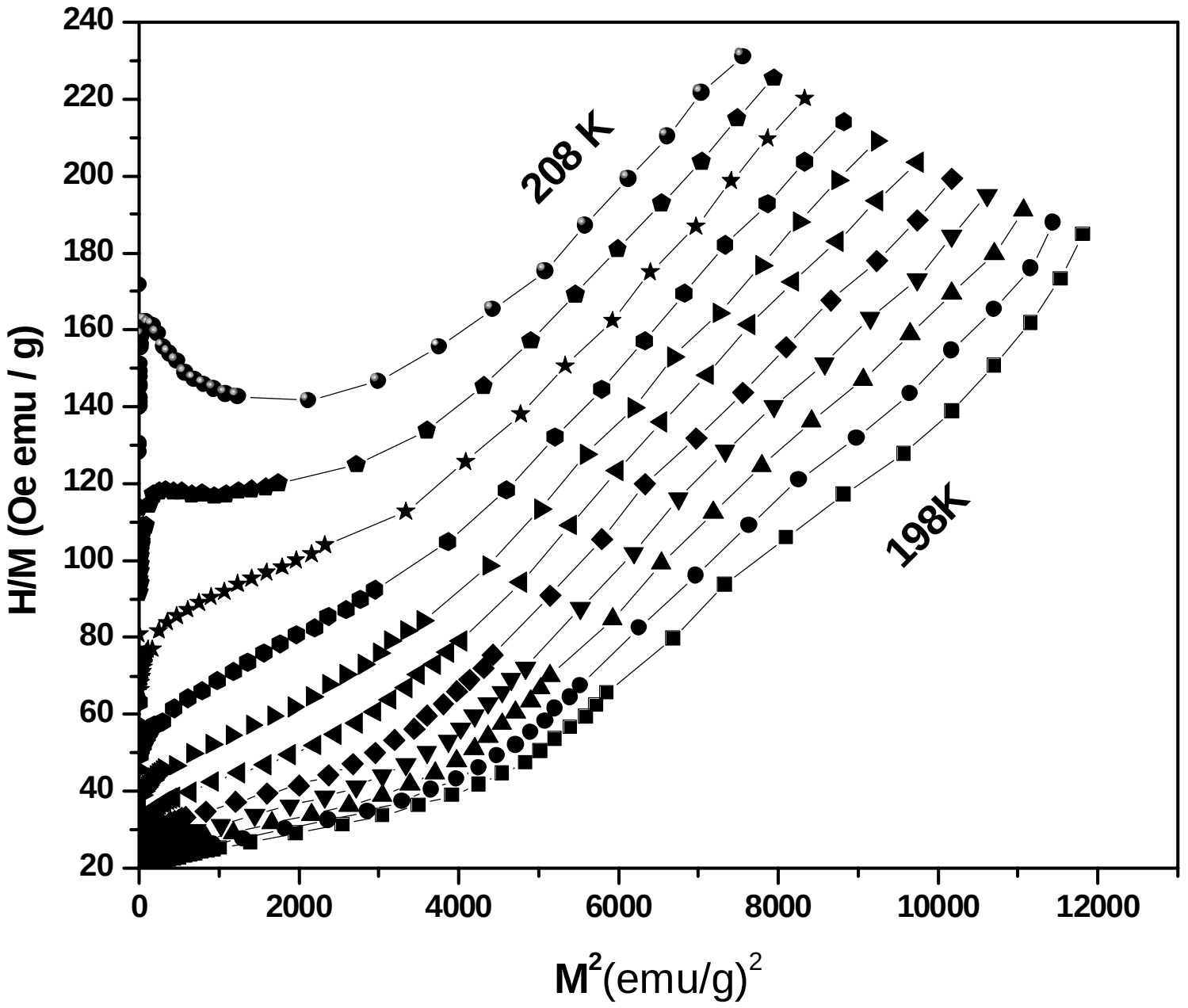

Figure 7 (b)

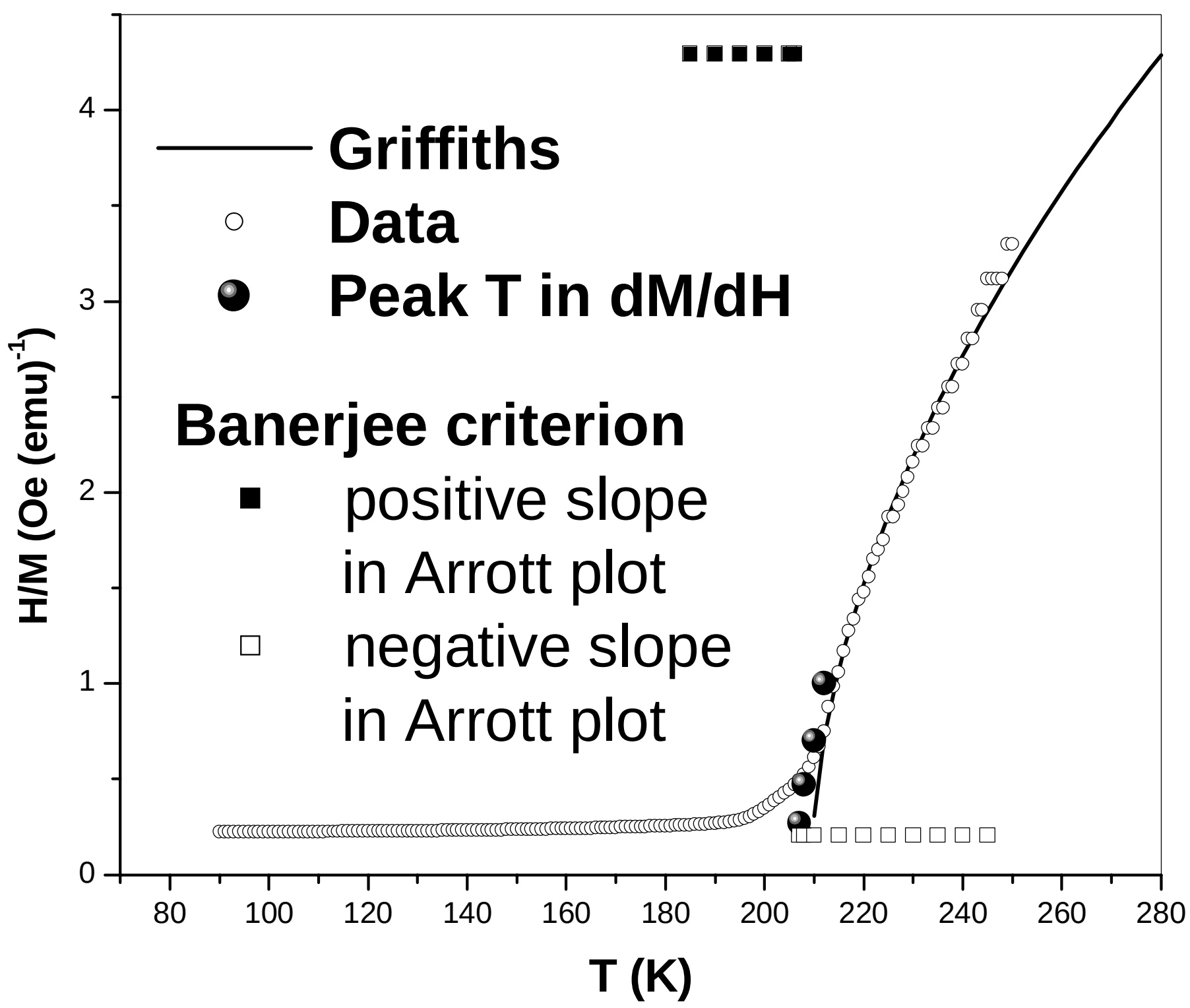

Figure 7 (c)

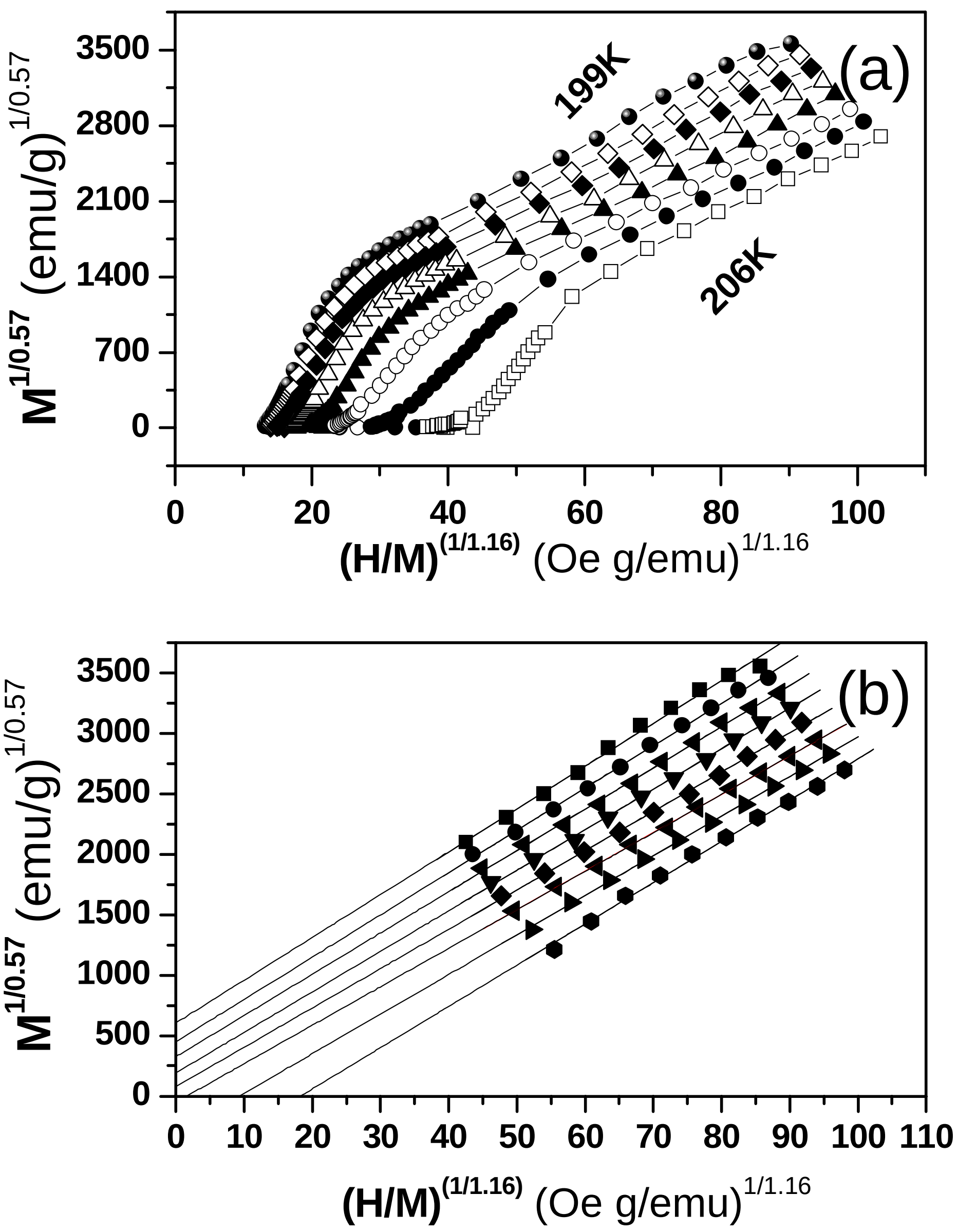

Figure 8 (a,b)

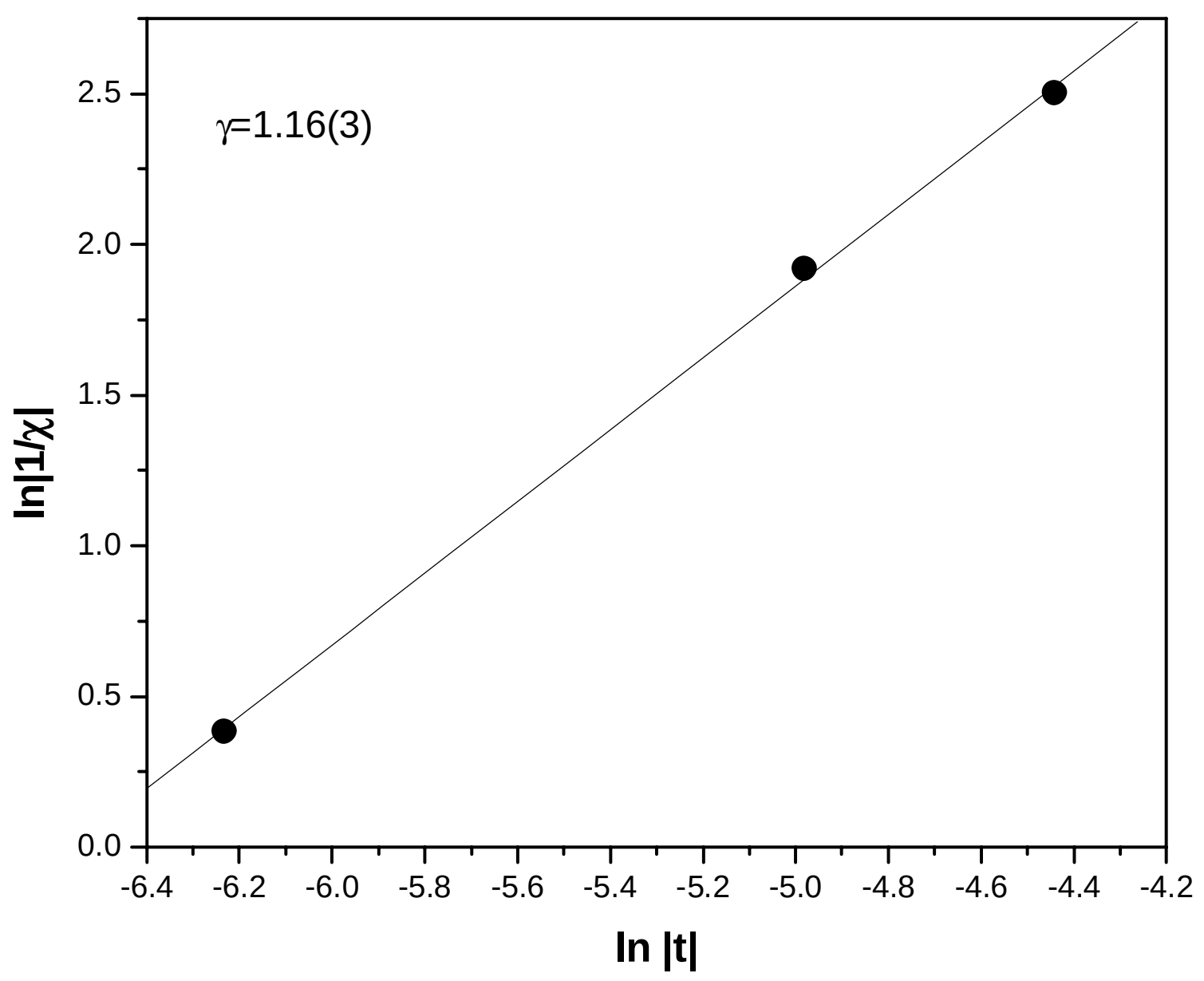

Figure 8(c)

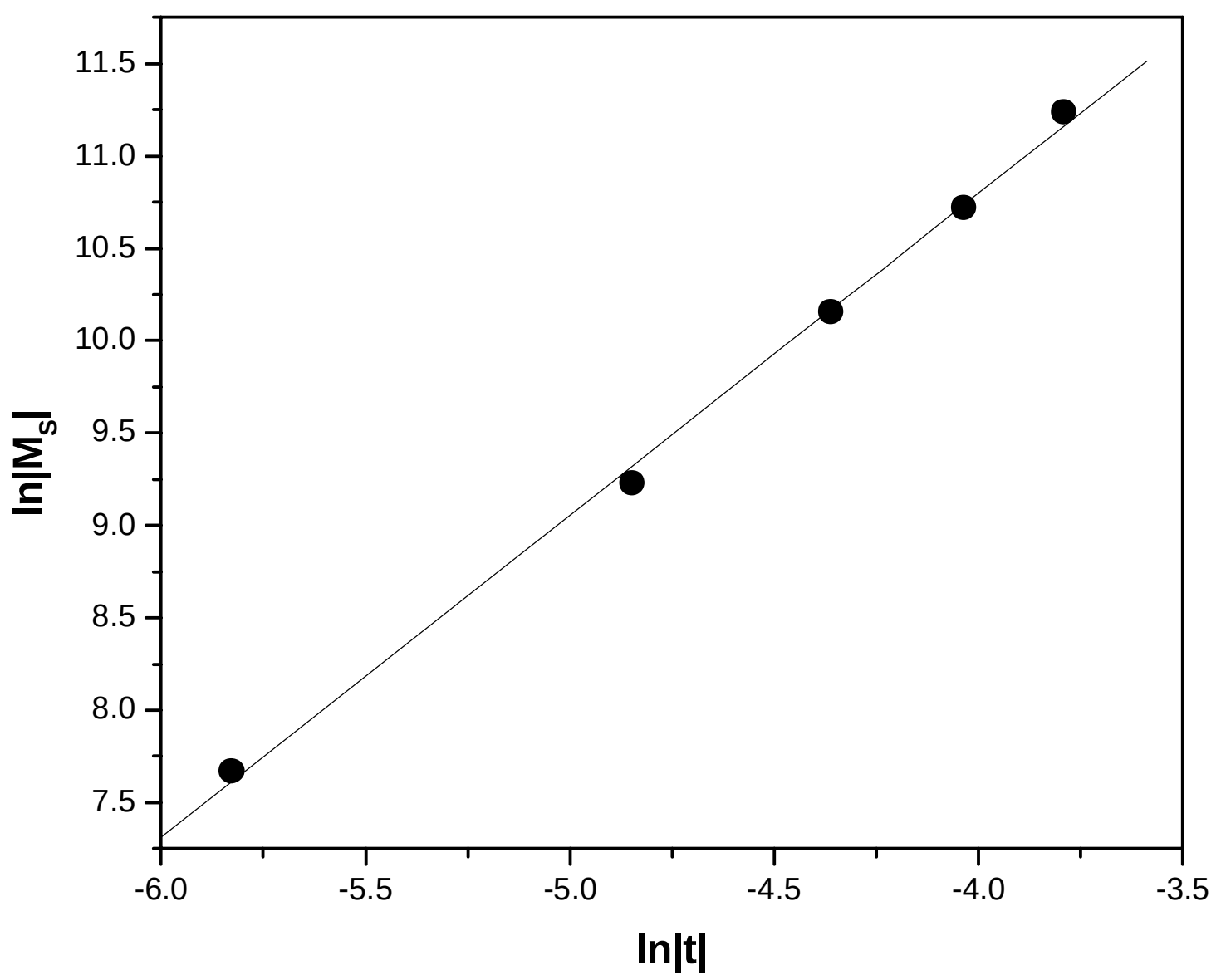

Figure 8(d)

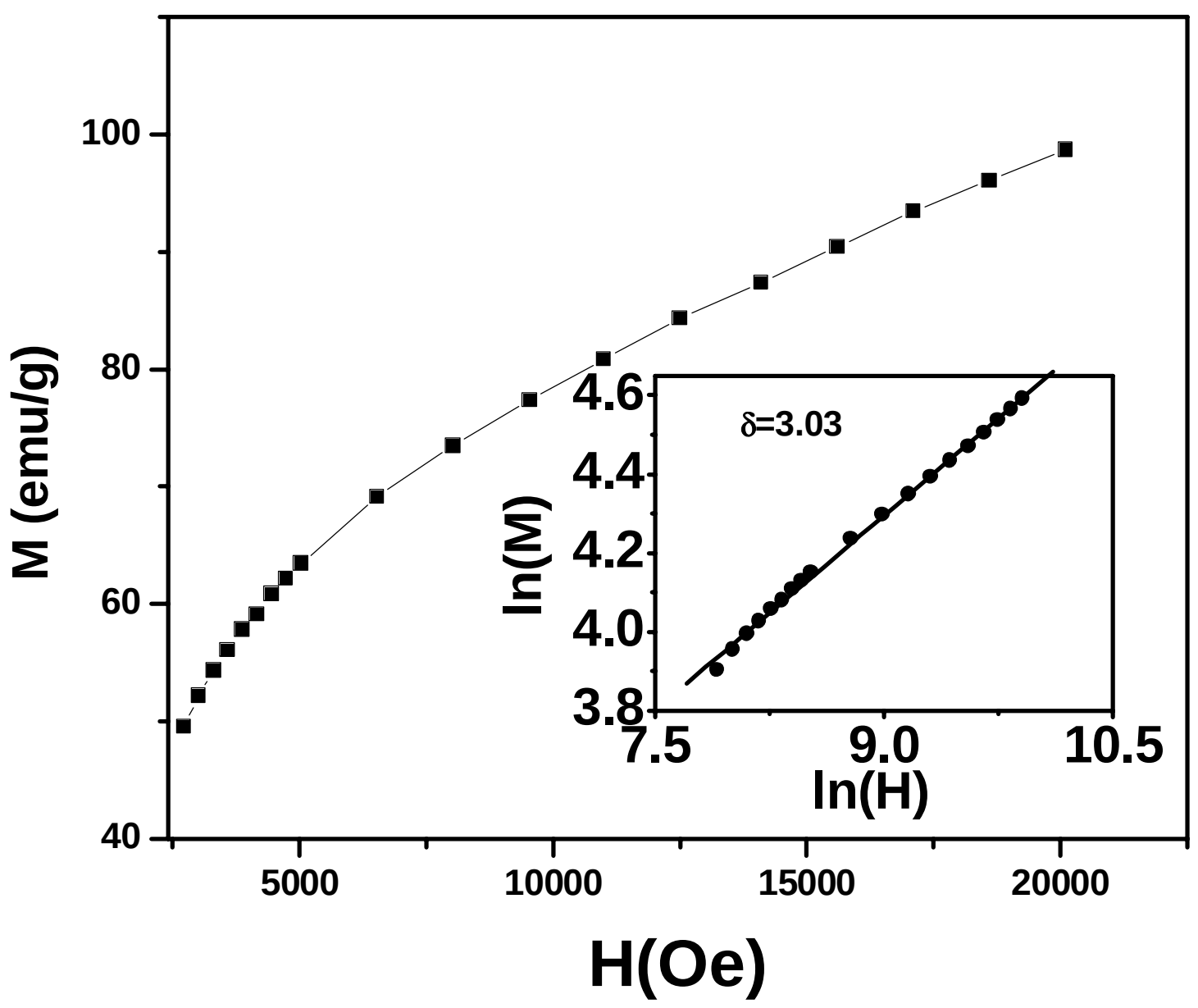

Figure 9

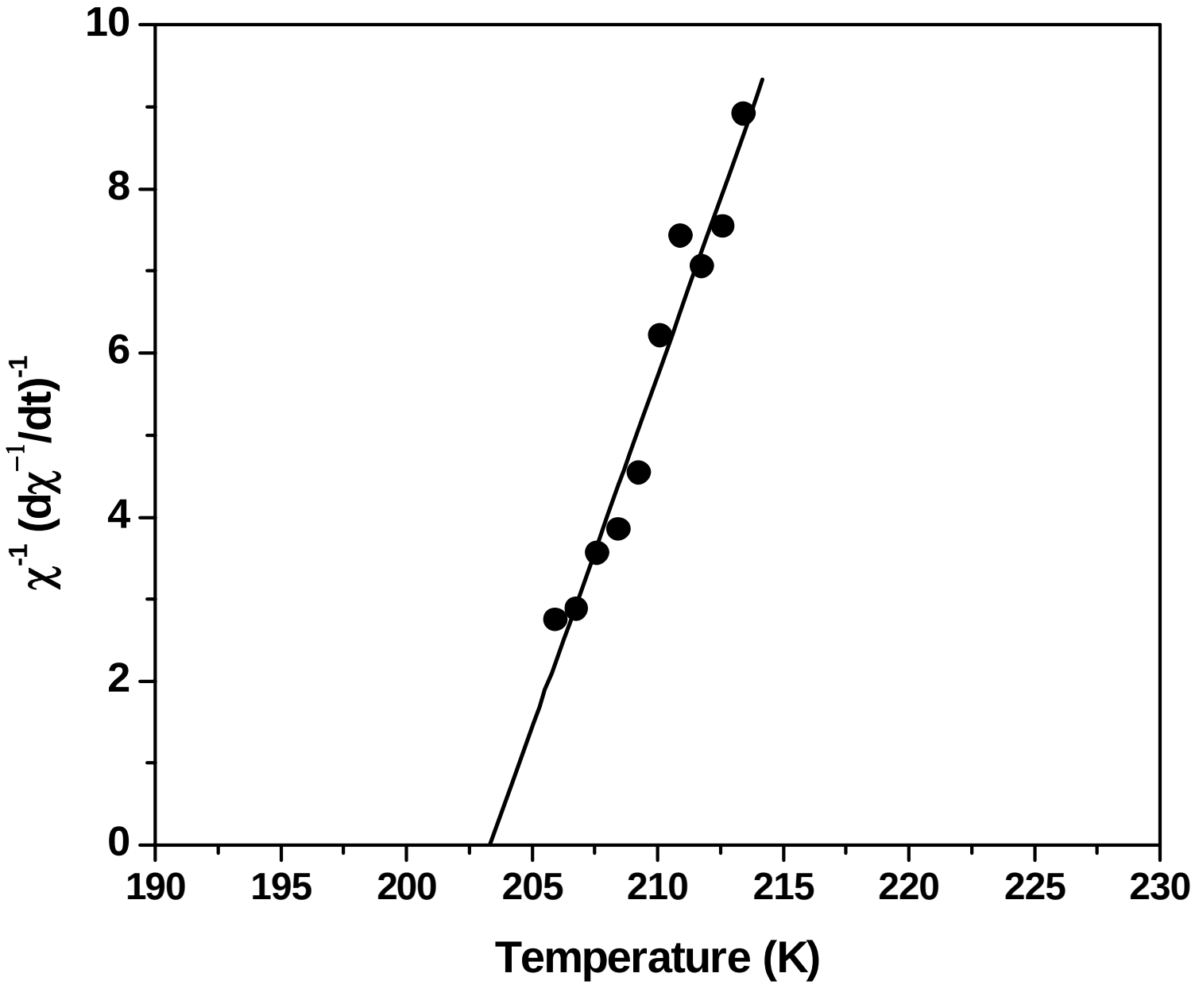

Figure 10

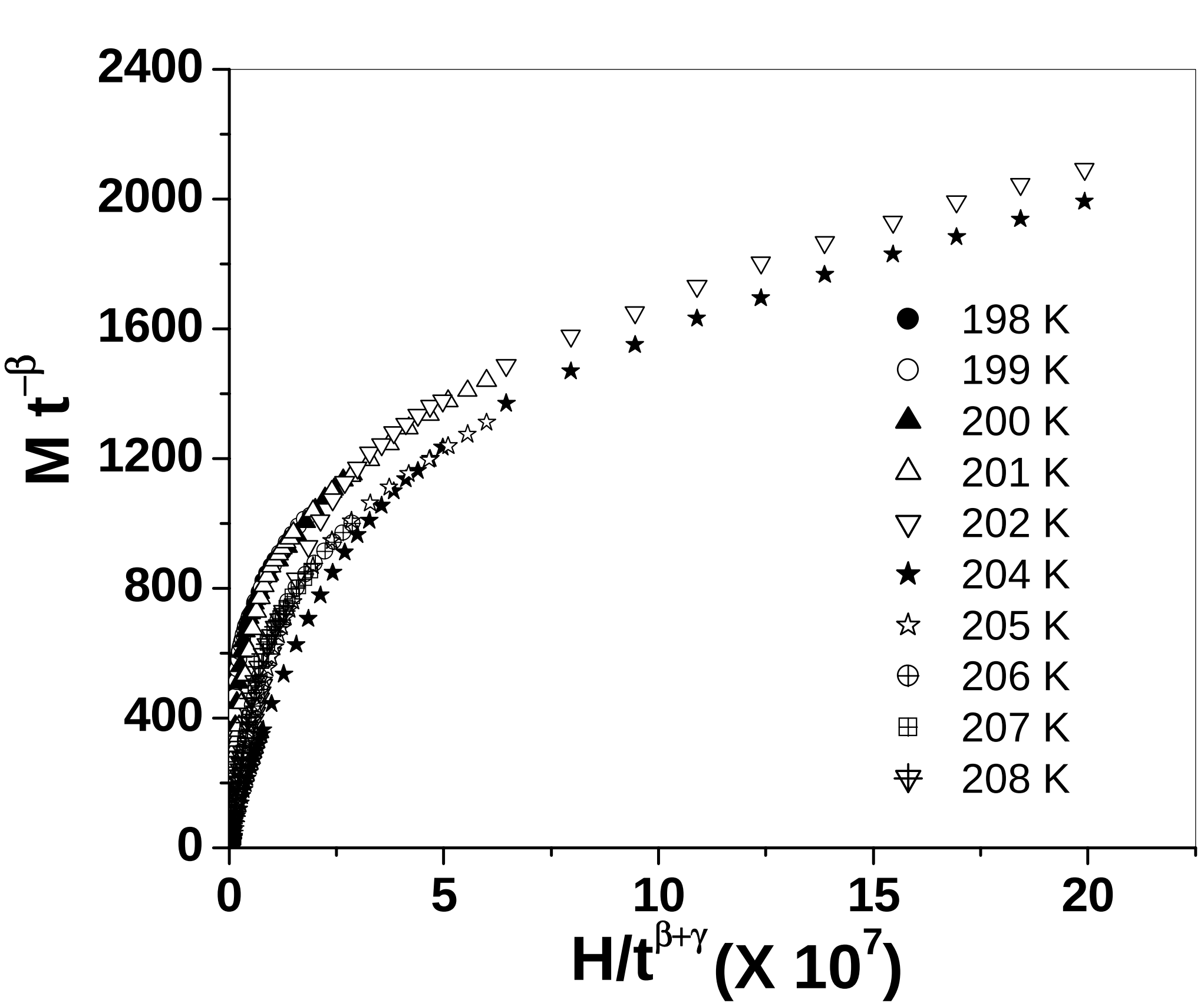

Fig 11 (a)

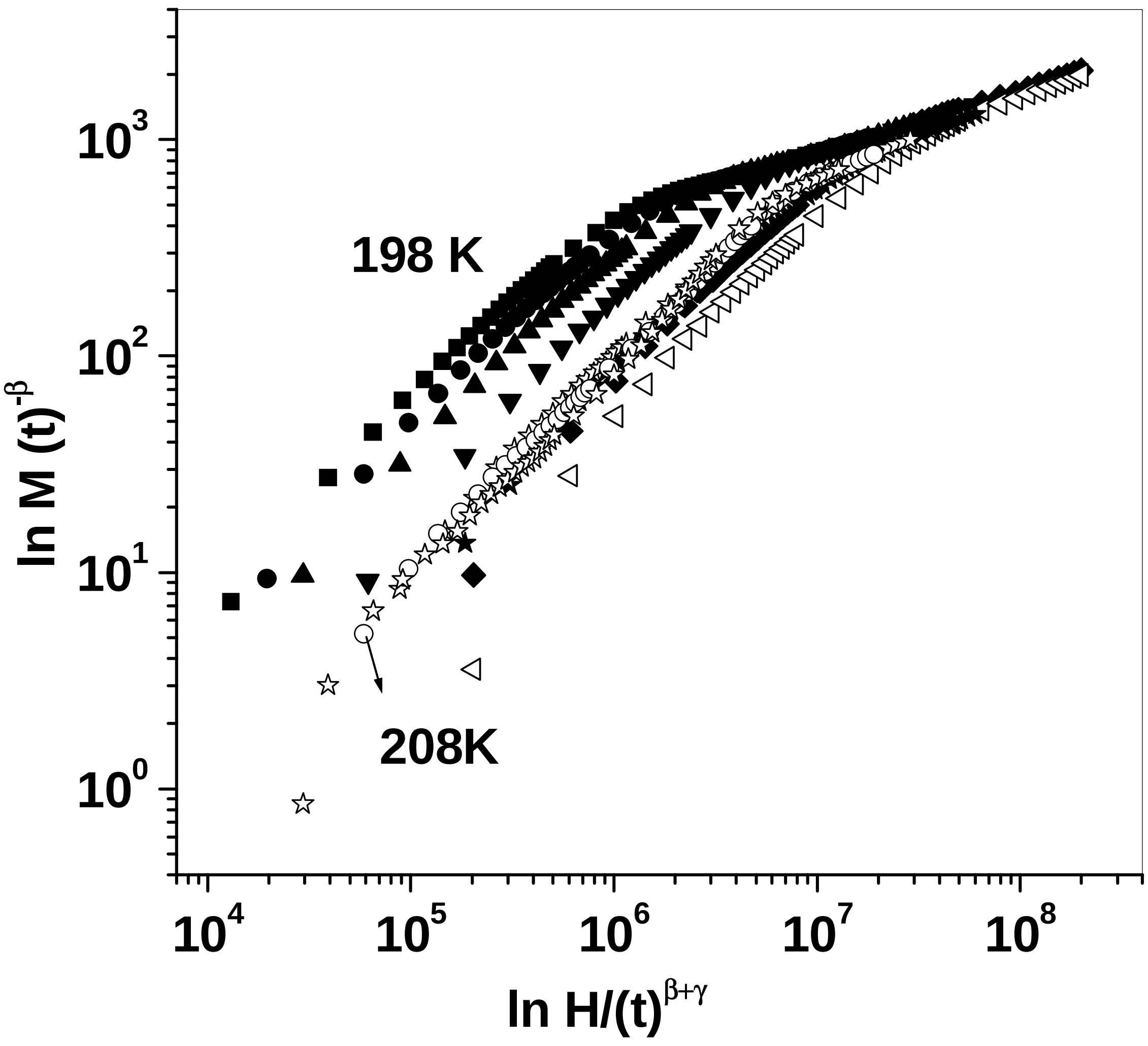

Fig 11 (b)

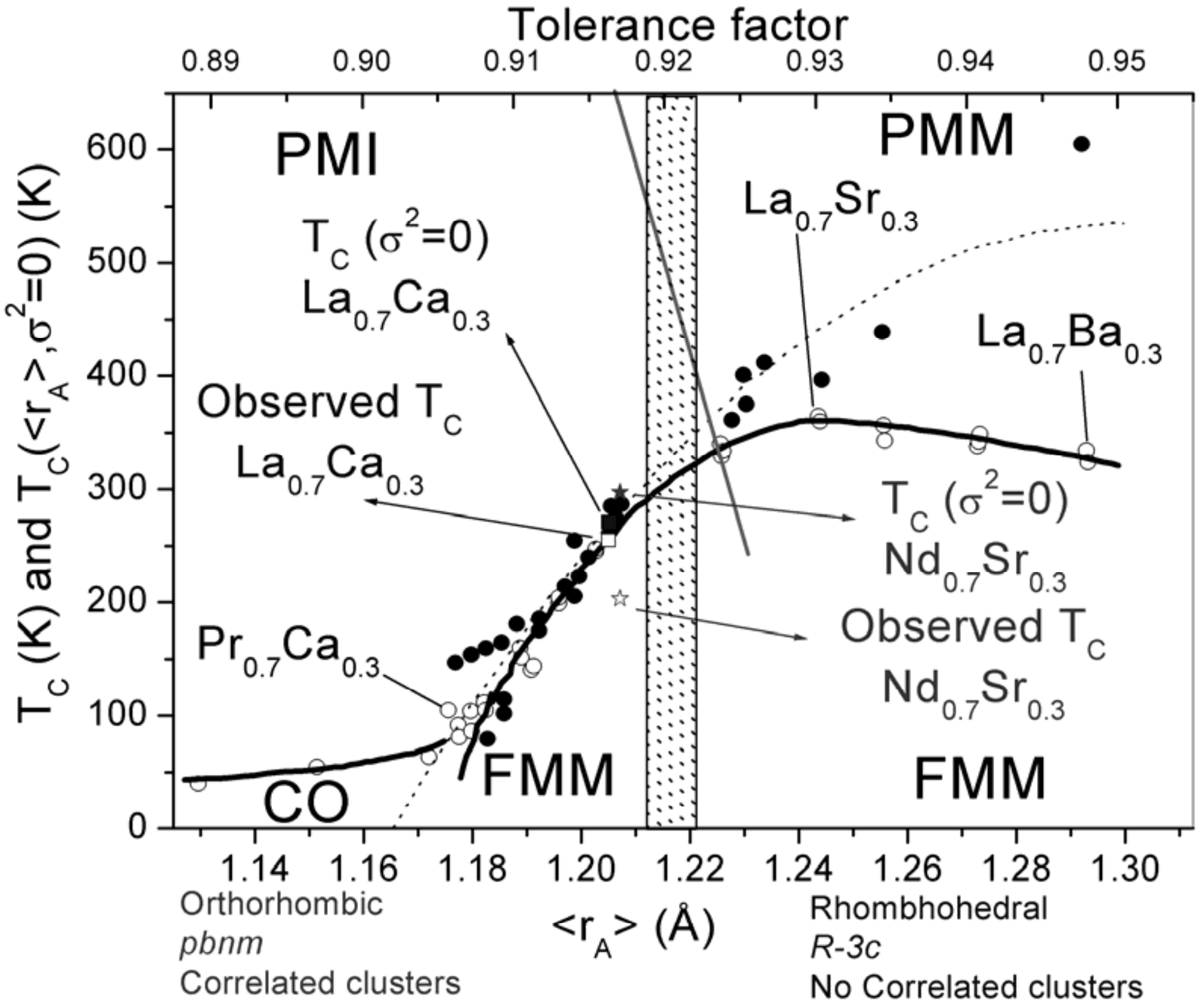

Figure 12